\documentclass[amsmath,aps,showpacs,a4paper,10pt]{revtex4}

 \usepackage{epsf}
 \usepackage{graphicx}    

\begin{document}

 \newcommand{\be}[1]{\begin{equation}\label{#1}}
 \newcommand{\ee}{\end{equation}}
 \newcommand{\bea}{\begin{eqnarray}}
 \newcommand{\eea}{\end{eqnarray}}
 \def\disp{\displaystyle}

 \def\gsim{ \lower .75ex \hbox{$\sim$} \llap{\raise .27ex \hbox{$>$}} }
 \def\lsim{ \lower .75ex \hbox{$\sim$} \llap{\raise .27ex \hbox{$<$}} }

 \begin{titlepage}

 \begin{flushright}
 arXiv:0812.4489
 \end{flushright}

 \title{\Large \bf Relaxing the Cosmological Constraints on
 Unparticle~Dark~Component}

 \author{Hao~Wei\,}
 \email[\,email address:\ ]{haowei@bit.edu.cn}
 \affiliation{Department of Physics, Beijing Institute
 of Technology, Beijing 100081, China}

 \begin{abstract}\vspace{1cm}
 \centerline{\bf ABSTRACT}\vspace{2mm}
Unparticle physics has been an active field since the seminal
 work of Georgi. Recently, many constraints on unparticles
 from various observations have been considered in the literature.
 In particular, the cosmological constraints on the unparticle
 dark component put it in a serious situation. In this
 work, we try to find a way out of this serious situation, by
 including the possible interaction between dark energy and the
 unparticle dark component.
 \end{abstract}


 \maketitle

 \end{titlepage}

 \renewcommand{\baselinestretch}{1.6}



\section{Introduction}\label{sec1}
Recently, the so-called unparticle physics has been an active
 field since the seminal work of Georgi~\cite{r1,r2}. It was
 based on the hypothesis that there could be an exact scale
 invariant hidden sector resisted at a high energy scale. A
 prototype model of such a sector is given by the Banks-Zaks
 theory which flows to an infrared fixed point at a lower
 energy scale $\Lambda_{\rm u}$ through dimensional
 transmutation~\cite{r3}. Recently, there has been a flood of
 papers on the unparticle phenomenology. We refer to
 e.g.~\cite{r3,r4} for some brief reviews. In fact, soon
 after the seminal work of Georgi~\cite{r1,r2}, some authors
 considered the constraints on unparticles from various
 observations, such as the collider experiments~\cite{r5}, the
 new long range force experiments~\cite{r6,r7}, the solar and
 reactor neutrinos data~\cite{r8}, as well as the observations
 from astrophysics and cosmology~\cite{r7,r9,r10,r11,r23}.
 These observations put fairly stringent constraints on the
 unparticle.

The unparticle does not have a definite mass and instead has a
 continuous spectral density as a consequence of scale
 invariance~\cite{r1,r2} (see also e.g.~\cite{r11,r12,r24}),
 \be{eq1}
 \rho(P^2)=A_{d_{\rm u}}\theta(P^0)\theta(P^2)(P^2)^{d_{\rm u}-2},
 \ee
 where $P$ is the 4-momentum; $A_{d_{\rm u}}$ is the normalization
 factor; $d_{\rm u}$ is the scaling dimension. When $d_{\rm u}=1$,
 Eq.~(\ref{eq1}) reduces to the familiar one of a massless particle.
 The scaling dimension $d_{\rm u}$ is constrained by the unitarity
 of the conformal algebra and the requirement to avoid the singular
 behavior~\cite{r1,r4}. The theoretical bounds are
 $1\leq d_{\rm u}\leq 2$ (for bosonic unparticles) or
 $3/2\leq d_{\rm u}\leq 5/2$ (for fermionic
 unparticles)~\cite{r4}. Typically, values $1\leq d_{\rm u}\leq 2$
 have been extensively considered in the literature.

The pressure and energy density of thermal unparticles have
 been derived in~\cite{r12}. They are
 \bea
 &&p_{\rm u}=g_s T^4\left(\frac{T}{\Lambda_{\rm u}}
 \right)^{2\left(d_{\rm u}-1\right)}\frac{{\cal C}
 \left(d_{\rm u}\right)}{4\pi^2}\,,\label{eq2}\\
 &&\rho_{\rm u}=\left(2d_{\rm u}+1\right)g_s T^4\left(\frac{T}
 {\Lambda_{\rm u}}\right)^{2\left(d_{\rm u}-1\right)}
 \frac{{\cal C}\left(d_{\rm u}\right)}{4\pi^2}\,,\label{eq3}
 \eea
 where ${\cal C}\left(d_{\rm u}\right)=B\left(3/2,d_{\rm u}\right)
 \Gamma\left(2d_{\rm u}+2\right)\zeta\left(2d_{\rm u}+2\right)$,
 while $B$, $\Gamma$, $\zeta$ are the Beta, Gamma and Zeta
 functions, respectively. These are the results for the bosonic
 unparticles. The pressure and energy density of fermionic
 unparticles can be obtained by replacing
 ${\cal C}\left(d_{\rm u}\right)$ by $\left[1-2^{-\left(2d_{\rm u}
 +1\right)}\right]{\cal C}\left(d_{\rm u}\right)$. Therefore,
 the equation-of-state parameter (EoS) of both the bosonic and
 fermionic unparticles is given by~\cite{r12}
 \be{eq4}
 w_{\rm u}\equiv\frac{p_{\rm u}}{\rho_{\rm u}}
 =\frac{1}{2d_{\rm u}+1}.
 \ee
 When $d_{\rm u}=1$, we have $w_{\rm u}=1/3$, which is the same
 as radiation. When $d_{\rm u}\to\infty$, we find $w_{\rm u}\to 0$,
 which approaches the EoS of pressureless matter. In the
 intermediate case, the EoS of unparticles is different from
 the one of radiation or cold dark matter, and generically lies
 in between. Since the unparticle interacts weakly with standard
 model particles, it is ``dark'' in this sense. Some authors
 regard the unparticle as a candidate of dark matter~\cite{r13}
 (see also~\cite{r11,r12}). Noting that the EoS of unparticles
 ($w_{\rm u}>0$) is different from the one of cold dark matter
 ($w_{\rm m}=0$), we instead call it ``dark component'' to avoid
 confusion.

In~\cite{r11}, the cosmological constraints on the unparticle dark
 component have been considered, by using the type Ia supernovae
 (SNIa), the shift parameter of the cosmic microwave background
 (CMB), and the baryon acoustic oscillation (BAO). The authors
 of~\cite{r11} found that $d_{\rm u}>60$ at $95\%$ confidence
 level (C.L.) for the $\Lambda$UDM model in which the unparticle is
 the sole dark matter. As mentioned above, however, the
 theoretical bounds on unparticles are $1\leq d_{\rm u}\leq 2$
 (for bosonic unparticles) or $3/2\leq d_{\rm u}\leq 5/2$ (for
 fermionic unparticles)~\cite{r4}. The situation is serious.
 Even for the $\Lambda$UCDM model in which the unparticle dark
 component co-exists with cold dark matter, they found that
 the unparticle dark component can at most make up a few
 percent of the total cosmic energy density if $d_{\rm u}<10$;
 so that it cannot be a major component~\cite{r11}. In fact,
 it is easy to understand these results. Since the unparticle dark
 component~scales as $a^{-3(1+w_{\rm u})}$ whereas cold dark
 matter scales as $a^{-3}$ (here $a$ is the scale factor of the
 universe), for $w_{\rm u}>0$, the energy density of the unparticle
 dark component decreases faster than the one of cold dark matter.
 So, it is not surprising that the energy density of the unparticle
 dark component cannot be comparable with cold dark matter at
 the present epoch. To make a considerable contribution to the
 total cosmic energy density, $d_{\rm u}$ should be very large
 to make $w_{\rm u}\simeq 0$ so that the unparticle dark component
 could mimic cold dark matter.

In this work, we try to find a way out of the serious
 situation mentioned above. Our physical motivation is very
 simple. If dark energy, the major component of the universe,
 can decay into unparticles, the energy density of the unparticle
 dark component should decrease {\em more slowly}. Since both dark
 energy and unparticle are unseen, the interaction between
 them is not prevented. If the dilution rates of unparticle
 dark component and cold dark matter are comparable, it is
 possible to have a considerable energy density of the unparticle
 dark component at the present epoch, without requiring
 $d_{\rm u}$ to be very large. In fact, this is similar to the
 key point of the interacting dark energy models which are
 extensively considered in the literature to alleviate the
 cosmological coincidence problem (see e.g.~\cite{r14,r25} and
 references therein).

Here, we consider a flat universe which contains dark energy,
 the unparticle dark component and pressureless matter (including
 cold dark matter and baryons). We assume that dark energy and
 unparticle dark component exchange energy according to (see
 e.g.~\cite{r14,r25} and references therein)
 \bea
 &&\dot{\rho}_{\rm de}+3H\rho_{\rm de}(1+w_{\rm de})=-Q,\label{eq5}\\
 &&\dot{\rho}_{\rm u}+3H\rho_{\rm u}(1+w_{\rm u})=Q,\label{eq6}
 \eea
 whereas pressureless matter (including cold dark matter and
 baryons) evolves independently, i.e.,
 \be{eq7}
 \dot{\rho}_{\rm m}+3H\rho_{\rm m}=0.
 \ee
 So, the total energy conservation equation
 $\dot{\rho}_{\rm tot}+3H\rho_{\rm tot}(1+w_{\rm tot})=0$
 is preserved. $H\equiv\dot{a}/a$ is the Hubble parameter;
 $a=(1+z)^{-1}$ is the scale factor of the universe (we set
 $a_0=1$); the subscript ``0'' indicates the present value
 of the corresponding quantity; $z$ is the redshift; a dot
 denotes the derivative with respect to cosmic time $t$.
 The interaction forms extensively considered in the literature
 (see for instance~\cite{r14,r25} and references therein) are
 $Q\propto H\rho_{\rm m}$, $H\rho_{\rm de}$, $H\rho_{\rm tot}$,
 $\kappa\rho_{\rm m}\dot{\phi}$ (where $\kappa^2\equiv 8\pi G$),
 and so on. In this work, for simplicity, we consider the
 interaction term
 \be{eq8}
 Q=3\alpha H\rho_{\rm u},
 \ee
 where $\alpha$ is a constant. So, Eq.~(\ref{eq6}) becomes
 $\dot{\rho}_{\rm u}+3H\rho_{\rm u}\left(1+w_{\rm u}^{\rm
 eff}\right)=0$, where $w_{\rm u}^{\rm eff}\equiv w_{\rm u}-\alpha$.
 It is easy to find that
 \be{eq9}
 \rho_{\rm u}=\rho_{\rm u0}\,a^{-3\left(1+
 w_{\rm u}^{\rm eff}\right)}.
 \ee
 On the other hand, from Eq.~(\ref{eq7}), we have
 $\rho_{\rm m}=\rho_{\rm m0}\,a^{-3}$. For convenience, we
 introduce the fractional energy density
 $\Omega_i\equiv\left(8\pi G\rho_i\right)/\left(3H^2\right)$,
 where $i=\rm de$, u and m.

In the following sections, similar to~\cite{r11}, we consider
 the cosmological constraints on the unparticle dark component
 which interacts with dark energy, by using the observations
 of SNIa, the shift parameter $R$ of CMB, and the distance
 parameter $A$ from BAO. In Sec.~\ref{sec2}, we present the
 cosmological data and the methodology used to constrain the
 models. In Sec.~\ref{sec3} and \ref{sec4}, we consider the
 observational constraints on the interacting $\Lambda$UCDM
 model and the interacting XUCDM model, respectively. As
 expected, we find that the cosmological constraints on the
 unparticle dark component can be significantly relaxed, thanks
 to the possible interaction between dark energy and unparticle
 dark component. The tension between theoretical bounds and
 cosmological constraints could be removed. Finally, some
 concluding remarks are given in Sec.~\ref{sec5}.

\vspace{-2mm}  


\section{Cosmological data}\label{sec2}
In this work, we perform a $\chi^2$ analysis to obtain the
 constraints on the model parameters. The data points of the
 307 Union SNIa compiled in~\cite{r15} are given in terms
 of the distance modulus $\mu_{obs}(z_i)$. On the other hand,
 the theoretical distance modulus is defined as
 \be{eq10}
 \mu_{th}(z_i)\equiv 5\log_{10}D_L(z_i)+\mu_0,
 \ee
 where $\mu_0\equiv 42.38-5\log_{10}h$ and $h$ is the Hubble
 constant $H_0$ in units of $100~{\rm km/s/Mpc}$, whereas
 \be{eq11}
 D_L(z)=(1+z)\int_0^z \frac{d\tilde{z}}{E(\tilde{z};{\bf p})},
 \ee
 in which $E\equiv H/H_0$; ${\bf p}$ denotes the model parameters.
 The $\chi^2$ from the 307 Union SNIa are given by
 \be{eq12}
 \chi^2_{\mu}({\bf p})=\sum\limits_{i}
 \frac{\left[\mu_{obs}(z_i)-\mu_{th}(z_i)\right]^2}{\sigma^2(z_i)},
 \ee
 where $\sigma$ is the corresponding $1\sigma$ error. The parameter
 $\mu_0$ is a nuisance parameter but it is independent of the data
 points. One can perform an uniform marginalization over $\mu_0$.
 However, there is an alternative way. Following~\cite{r16,r17}, the
 minimization with respect to $\mu_0$ can be made by expanding the
 $\chi^2_{\mu}$ of Eq.~(\ref{eq12}) with respect to $\mu_0$ as
 \be{eq13}
 \chi^2_{\mu}({\bf p})=\tilde{A}-2\mu_0\tilde{B}+\mu_0^2\tilde{C},
 \ee
 where
 $$\tilde{A}({\bf p})=\sum\limits_{i}\frac{\left[\mu_{obs}(z_i)
 -\mu_{th}(z_i;\mu_0=0,{\bf p})\right]^2}
 {\sigma_{\mu_{obs}}^2(z_i)}\,,$$
 $$\tilde{B}({\bf p})=\sum\limits_{i}\frac{\mu_{obs}(z_i)
 -\mu_{th}(z_i;\mu_0=0,{\bf p})}{\sigma_{\mu_{obs}}^2(z_i)}\,,
 ~~~~~~~~~~
 \tilde{C}=\sum\limits_{i}\frac{1}{\sigma_{\mu_{obs}}^2(z_i)}\,.$$
 Eq.~(\ref{eq13}) has a minimum for
 $\mu_0=\tilde{B}/\tilde{C}$ at
 \be{eq14}
 \tilde{\chi}^2_{\mu}({\bf p})=
 \tilde{A}({\bf p})-\frac{\tilde{B}({\bf p})^2}{\tilde{C}}.
 \ee
 Since $\chi^2_{\mu,\,min}=\tilde{\chi}^2_{\mu,\,min}$
 obviously, we can instead minimize $\tilde{\chi}^2_{\mu}$
 which is independent of $\mu_0$. The shift parameter $R$ of
 CMB is defined by~\cite{r18,r19}
 \be{eq15}
 R\equiv\Omega_{\rm m0}^{1/2}\int_0^{z_\ast}
 \frac{d\tilde{z}}{E(\tilde{z})},
 \ee
 where the redshift of recombination $z_\ast=1090$ which has
 been updated in the Wilkinson Microwave Anisotropy Probe
 5-year (WMAP5) data~\cite{r20}. The shift parameter $R$
 relates the angular diameter distance to the last scattering
 surface, the comoving size of the sound horizon at $z_\ast$
 and the angular scale of the first acoustic peak in CMB power
 spectrum of temperature fluctuations~\cite{r18,r19}. The value
 of $R$ has been updated to $1.710\pm 0.019$ from the WMAP5
 data~\cite{r20}. On the other hand, the distance parameter $A$
 of the measurement of the BAO peak in the distribution of SDSS
 luminous red galaxies~\cite{r21} is given by
 \be{eq16}
 A\equiv\Omega_{\rm m0}^{1/2}E(z_b)^{-1/3}\left[\frac{1}{z_b}
 \int_0^{z_b}\frac{d\tilde{z}}{E(\tilde{z})}\right]^{2/3},
 \ee
 where $z_b=0.35$. In~\cite{r22}, the value of $A$ has been
 determined to be $0.469\,(n_s/0.98)^{-0.35}\pm 0.017$. Here
 the scalar spectral index $n_s$ is taken to be $0.960$, which
 has been updated from the WMAP5 data~\cite{r20}. So, the total
 $\chi^2$ is given by
 \be{eq17}
 \chi^2=\tilde{\chi}^2_{\mu}+\chi^2_{CMB}+\chi^2_{BAO},
 \ee
 where $\tilde{\chi}^2_{\mu}$ is given in Eq.~(\ref{eq14}),
 $\chi^2_{CMB}=(R-R_{obs})^2/\sigma_R^2$ and
 $\chi^2_{BAO}=(A-A_{obs})^2/\sigma_A^2$. The best-fit model
 parameters are determined by minimizing the total $\chi^2$.
 As is well known, the $68\%$ C.L. is determined by
 $\Delta\chi^2\equiv\chi^2-\chi^2_{min}\leq 1.00$, $2.30$,
 $3.53$, $4.72$ and $5.89$ for $n=1$, $2$, $3$, $4$ and $5$,
 respectively, where $n$ is the number of free model
 parameters. Similarly, the $95\%$ C.L. is determined by
 $\Delta\chi^2\equiv\chi^2-\chi^2_{min}\leq 4.00$, $6.17$,
 $8.02$, $9.70$ and $11.3$ for $n=1$, $2$, $3$, $4$ and $5$,
 respectively.

In the following sections, we consider the observational
 constraints on the interacting $\Lambda$UCDM model and
 the interacting XUCDM model, respectively, by using the
 methodology presented here.


\section{Cosmological constraints on the
 interacting $\Lambda$UCDM model}\label{sec3}
In this section, we consider the interacting $\Lambda$UCDM
 model in which the role of dark energy is played by the
 decaying cosmological constant (vacuum energy), as
 in the well-known $\Lambda(t)$CDM model~\cite{r26,r27,r28}.
 In this case, Eq.~(\ref{eq5}) becomes
 \be{eq18}
 \dot{\rho}_\Lambda=-3\alpha H\rho_{\rm u}=
 -3\alpha\rho_{\rm u0}Ha^{-3\left(1+w_{\rm u}^{\rm eff}\right)}.
 \ee
 Noting that the case without interaction ($\alpha=0$) was
 considered in~\cite{r11}, there are two other different cases
 with $w_{\rm u}^{\rm eff}\not=-1$ and $w_{\rm u}^{\rm eff}=-1$
 for this differential equation. We will study them one by one.


 \begin{center}
 \begin{figure}[tbhp]
 \centering
 \includegraphics[width=0.98\textwidth]{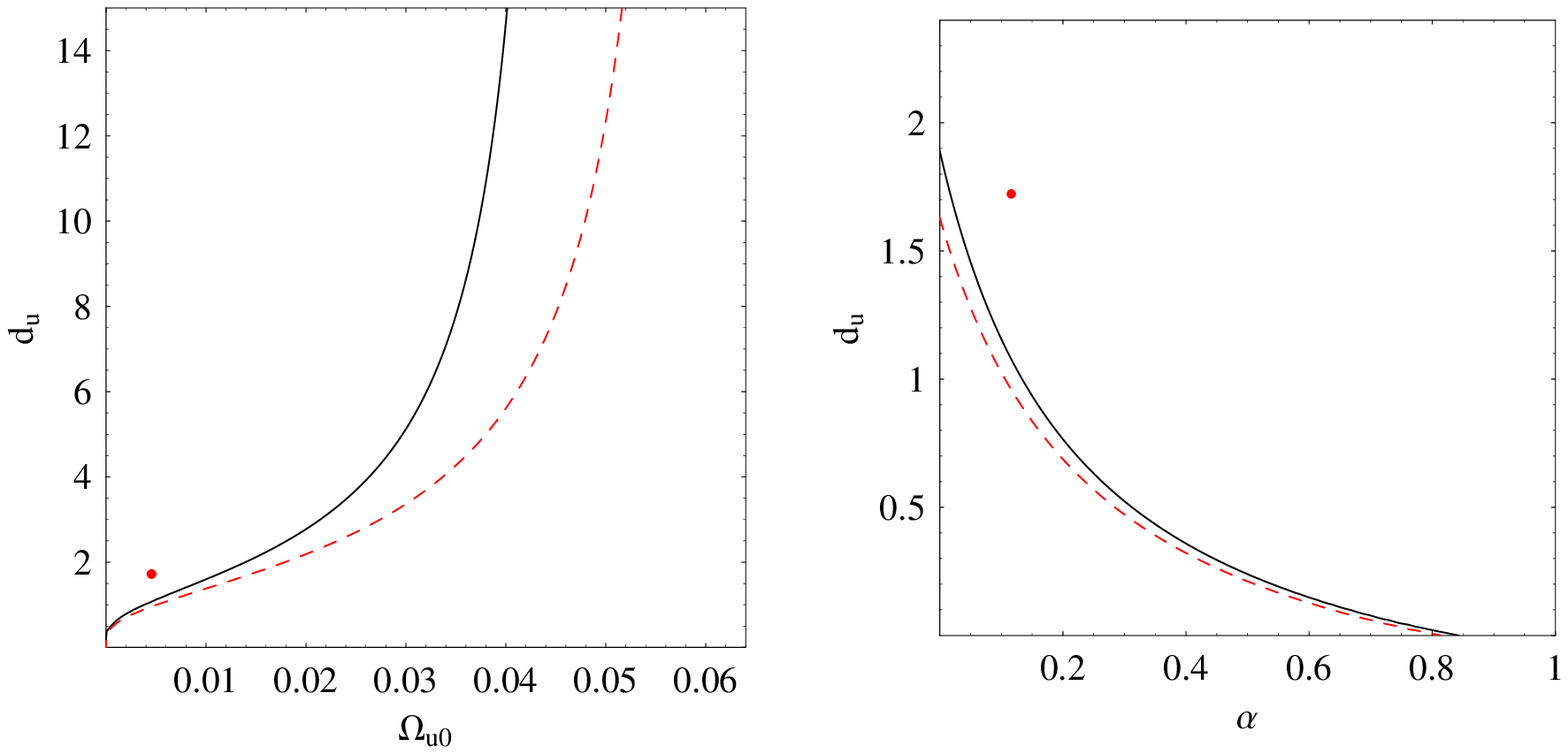}
 \caption{\label{fig1}
 The $68\%$ C.L. (black solid lines) and $95\%$ C.L. (red
 dashed lines) contours in the $\Omega_{\rm u0}-d_{\rm u}$
 plane and the $\alpha-d_{\rm u}$ plane for the interacting
 $\Lambda$UCDM model with $w_{\rm u}^{\rm eff}\not=-1$. The
 best fit is indicated by the red solid point.}
 \end{figure}
 \end{center}


\vspace{-10mm}  


\subsection{The interacting $\Lambda$UCDM model
 with $w_{\rm u}^{\rm eff}\not=-1$}\label{sec3a}
If $w_{\rm u}^{\rm eff}\not=-1$, the solution for Eq.~(\ref{eq18})
 is given by
 \be{eq19}
 \rho_\Lambda=\frac{\alpha\rho_{\rm u0}}{1+w_{\rm u}^{\rm eff}}
 \,a^{-3\left(1+w_{\rm u}^{\rm eff}\right)}+const.,
 \ee
 where $const.$ is an integration constant. Obviously, this
 solution diverges for $w_{\rm u}^{\rm eff}=-1$ which
 will be considered in the next subsection. Inserting
 Eq.~(\ref{eq19}) into the Friedmann equation
 $H^2=\kappa^2\left(\rho_\Lambda+\rho_{\rm u}+\rho_{\rm m}\right)/3$
 and requiring $E(z=0)=1$, one can determine the integration
 constant. Finally, we find that
 \be{eq20}
 E^2=\Omega_{\rm m0}(1+z)^3+\left(1+\frac{\alpha}{1+w_{\rm u}^{\rm
 eff}}\right)\Omega_{\rm u0}(1+z)^{3\left(1+w_{\rm u}^{\rm
 eff}\right)}+\left[1-\Omega_{\rm m0}-\left(1+\frac{\alpha}{1+
 w_{\rm u}^{\rm eff}}\right)\Omega_{\rm u0}\right].
 \ee
 There are 4 free model parameters, namely, $\Omega_{\rm m0}$,
 $\Omega_{\rm u0}$, $d_{\rm u}$ and $\alpha$. By minimizing the
 corresponding total $\chi^2$ in Eq.~(\ref{eq17}), we find the
 best-fit parameters $\Omega_{\rm m0}=0.280$,
 $\Omega_{\rm u0}=0.005$, $d_{\rm u}=1.722$ and $\alpha=0.116$,
 while $\chi^2_{min}=311.924$. In Fig.~\ref{fig1}, we present
 the corresponding $68\%$ and $95\%$ C.L. contours in the
 $\Omega_{\rm u0}-d_{\rm u}$ plane and the $\alpha-d_{\rm u}$
 plane for the interacting $\Lambda$UCDM model with
 $w_{\rm u}^{\rm eff}\not=-1$, while the other parameters are
 taken to be the corresponding best-fit values. Obviously, from
 the left panel of Fig.~\ref{fig1}, we see that the unparticle
 dark component can at most make up a few percent of the total
 cosmic energy density, so that it cannot be a major component
 in this case.


\subsection{The interacting $\Lambda$UCDM model
 with $w_{\rm u}^{\rm eff}=-1$}\label{sec3b}
\begin{center}
 \begin{figure}[tbhp]
 \centering
 \includegraphics[width=0.54\textwidth]{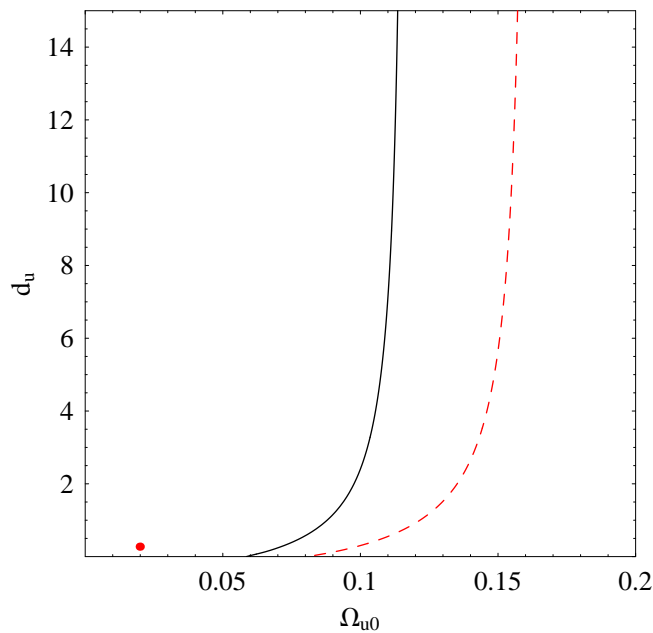}
 \caption{\label{fig2}
 The $68\%$ C.L. (black solid line) and $95\%$ C.L. (red
 dashed line) contours in the $\Omega_{\rm u0}-d_{\rm u}$
 plane for the interacting $\Lambda$UCDM model with
 $w_{\rm u}^{\rm eff}=-1$. The best fit is indicated by
 the red solid point.}
 \end{figure}
 \end{center}

As mentioned above, the solution Eq.~(\ref{eq19}) diverges
 for $w_{\rm u}^{\rm eff}=-1$. Therefore, one is required to
 consider this case separately. If $w_{\rm u}^{\rm eff}=-1$,
 we find that
 \be{eq21}
 \alpha=1+w_{\rm u}=1+\frac{1}{2d_{\rm u}+1}.
 \ee
 So, $\alpha$ is no longer a free model parameter. In this
 case, Eq.~(\ref{eq9}) becomes $\rho_{\rm u}=\rho_{\rm u0}$.
 Now, the solution for Eq.~(\ref{eq18}) is given by
 \be{eq22}
 \rho_\Lambda=-3\alpha\rho_{\rm u0}\ln a+const.,
 \ee
 where $const.$ is an integration constant. Then, inserting
 Eq.~(\ref{eq22}) into the Friedmann equation
 and requiring $E(z=0)=1$, one can determine the integration
 constant. Finally, we find that
 \be{eq23}
 E^2=\Omega_{\rm m0}(1+z)^3+3\Omega_{\rm u0}\left(1+
 \frac{1}{2d_{\rm u}+1}\right)\ln (1+z)+\left(1-
 \Omega_{\rm m0}\right).
 \ee
 In this case, there are 3 free model parameters, namely,
 $\Omega_{\rm m0}$, $\Omega_{\rm u0}$ and $d_{\rm u}$. By
 minimizing the corresponding total $\chi^2$ in
 Eq.~(\ref{eq17}), we find the best-fit parameters
 $\Omega_{\rm m0}=0.270$, $\Omega_{\rm u0}=0.020$ and
 $d_{\rm u}=0.273$, while $\chi^2_{min}=313.168$. In
 Fig.~\ref{fig2}, we present the corresponding $68\%$ and
 $95\%$ C.L. contours in the $\Omega_{\rm u0}-d_{\rm u}$
 plane for the interacting $\Lambda$UCDM model with
 $w_{\rm u}^{\rm eff}=-1$, while the other parameters are
 taken to be the corresponding best-fit values. It is a
 pleasure to find that the unparticle dark component can have a
 considerable contribution to the total cosmic energy density
 ($\Omega_{\rm u0}>0.1$) in the $95\%$ C.L. region while
 $1\leq d_{\rm u}\leq 2$ (for bosonic unparticles) or
 $3/2\leq d_{\rm u}\leq 5/2$ (for fermionic unparticles). The
 serious situation found in~\cite{r11} could be relaxed in
 this case.


\vspace{-11mm}  


\section{Cosmological constraints on the (interacting)
 XUCDM model}\label{sec4}
In this section, we consider the (interacting) XUCDM
 model in which the EoS of dark energy $w_{\rm x}$ is a
 constant. We can recast Eq.~(\ref{eq5}) as
 \be{eq24}
 \frac{d\rho_{\rm x}}{da}+\frac{3}{a}\,\rho_{\rm x}
 \left(1+w_{\rm x}\right)=-\frac{3\alpha}{a}\rho_{\rm u0}
 \,a^{-3\left(1+w_{\rm u}^{\rm eff}\right)}.
 \ee
 In the following, we first consider the case without
 interaction ($\alpha=0$). Then, we consider two different
 cases with $w_{\rm u}^{\rm eff}\not=w_{\rm x}$ and
 $w_{\rm u}^{\rm eff}=w_{\rm x}$ for the differential equation
 (\ref{eq24}).


\subsection{The XUCDM model without interaction}\label{sec4a}
\begin{center}
 \begin{figure}[tbhp]
 \centering
 \includegraphics[width=0.98\textwidth]{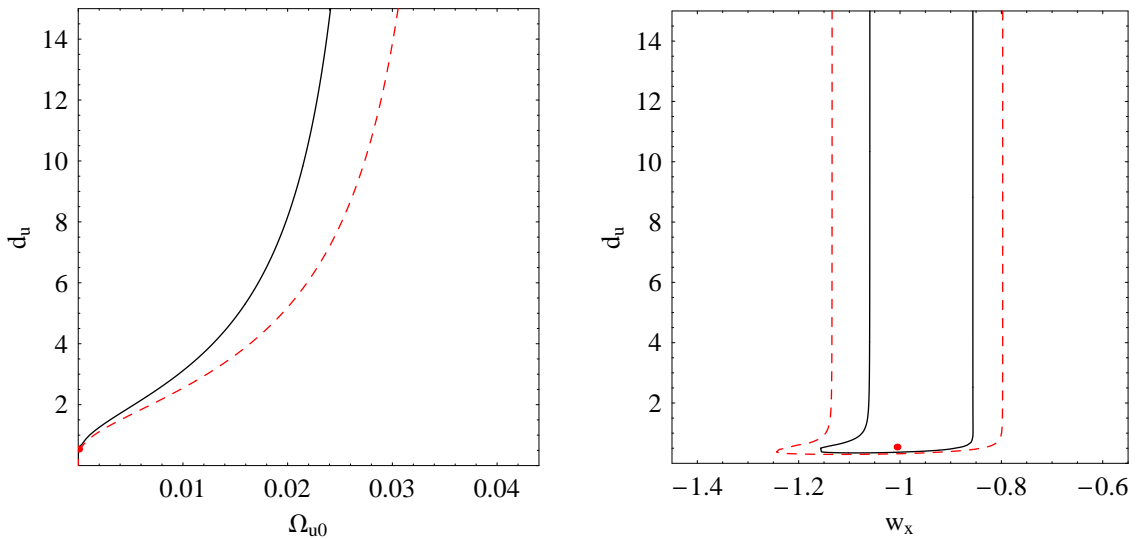}
 \caption{\label{fig3}
 The $68\%$ C.L. (black solid lines) and $95\%$ C.L. (red
 dashed lines) contours in the $\Omega_{\rm u0}-d_{\rm u}$
 plane and the $w_{\rm x}-d_{\rm u}$ plane for the XUCDM
 model without interaction. The best fit is indicated by
 the red solid point.}
 \end{figure}
 \end{center}

It is natural to see first what will happen if there is no
 interaction between dark energy and unparticle dark component.
 In this case, $\alpha=0$. One can easily write down
 \be{eq25}
 E^2=\Omega_{\rm m0}(1+z)^3+\Omega_{\rm u0}(1+z)^{3(1+
 w_{\rm u})}+\left(1-\Omega_{\rm m0}-\Omega_{\rm u0}\right)
 (1+z)^{3(1+w_{\rm x})}.
 \ee
 There are 4 free model parameters, namely, $\Omega_{\rm m0}$,
 $\Omega_{\rm u0}$, $d_{\rm u}$ and $w_{\rm x}$. By minimizing
 the corresponding total $\chi^2$ in Eq.~(\ref{eq17}), we find
 the best-fit parameters $\Omega_{\rm m0}=0.283$,
 $\Omega_{\rm u0}=0.0$, $d_{\rm u}=0.540$ and
 $w_{\rm x}=-1.006$, while $\chi^2_{min}=311.981$. In
 Fig.~\ref{fig3}, we present the corresponding $68\%$ and
 $95\%$ C.L. contours in the $\Omega_{\rm u0}-d_{\rm u}$ plane
 and the $w_{\rm x}-d_{\rm u}$ plane for the XUCDM model
 without interaction, while the other parameters are
 taken to be the corresponding best-fit values. Obviously,
 from the left panel of Fig.~\ref{fig3}, we see that the
 unparticle dark component can at most make up a few percent of
 the total cosmic energy density, so that it cannot be a major
 component in the case without interaction.


\vspace{-12mm}  


\subsection{The interacting XUCDM model with
 $w_{\rm u}^{\rm eff}\not=w_{\rm x}$}\label{sec4b}
\begin{center}
 \begin{figure}[tbhp]
 \centering
 \includegraphics[width=0.98\textwidth]{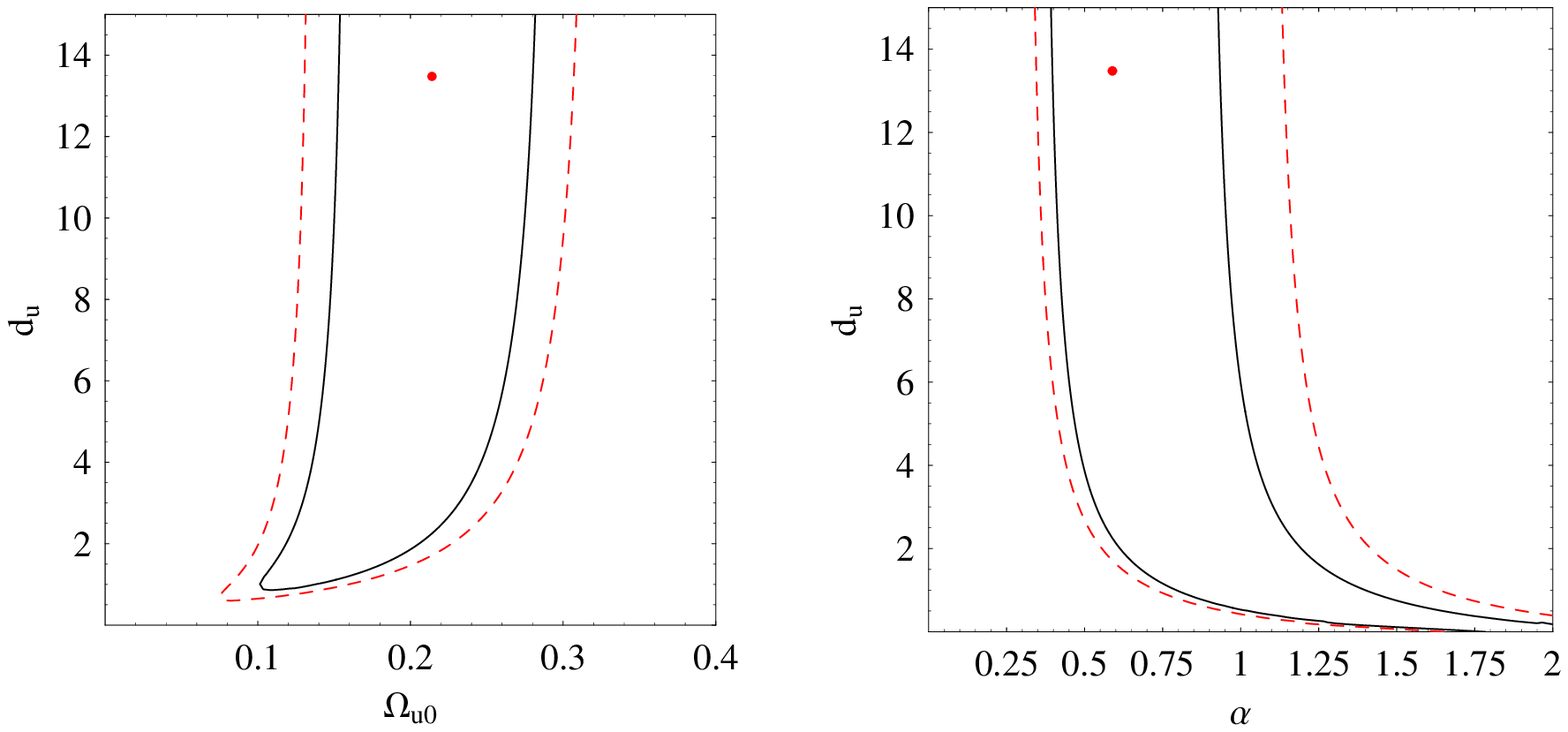}
 \caption{\label{fig4}
 The $68\%$ C.L. (black solid lines) and $95\%$ C.L. (red
 dashed lines) contours in the $\Omega_{\rm u0}-d_{\rm u}$
 plane and the $\alpha-d_{\rm u}$ plane for the interacting
 XUCDM model with $w_{\rm u}^{\rm eff}\not=w_{\rm x}$. The
 best fit is indicated by the red solid point.}
 \end{figure}
 \end{center}

If $w_{\rm u}^{\rm eff}\not=w_{\rm x}$, the solution for
 Eq.~(\ref{eq24}) is given by
 \be{eq26}
 \rho_{\rm x}=\frac{\alpha\rho_{\rm u0}}{w_{\rm u}^{\rm eff}
 -w_{\rm x}}\,a^{-3\left(1+w_{\rm u}^{\rm eff}\right)}+
 const.\times a^{-3\left(1+w_{\rm x}\right)},
 \ee
 where $const.$ is an integration constant. Obviously, this
 solution diverges for $w_{\rm u}^{\rm eff}=w_{\rm x}$ which
 will be considered in the next subsection. Inserting
 Eq.~(\ref{eq26}) into the Friedmann equation
 $H^2=\kappa^2\left(\rho_{\rm x}+\rho_{\rm u}+\rho_{\rm m}\right)/3$
 and requiring $E(z=0)=1$, one can determine the integration
 constant. Finally, we find that
 \bea
 E^2=\,\Omega_{\rm m0}(1+z)^3&+&\left(1+\frac{\alpha}{w_{\rm
 u}^{\rm eff}-w_{\rm x}}\right)\Omega_{\rm u0}(1+z)^{3\left(1+
 w_{\rm u}^{\rm eff}\right)}\nonumber\\
 &+&\left[1-\Omega_{\rm m0}-\left(1+\frac{\alpha}
 {w_{\rm u}^{\rm eff}-w_{\rm x}}\right)\Omega_{\rm u0}\right]
 (1+z)^{3\left(1+w_{\rm x}\right)}.\label{eq27}
 \eea
 There are 5 free model parameters, namely, $\Omega_{\rm m0}$,
 $\Omega_{\rm u0}$, $d_{\rm u}$, $\alpha$ and $w_{\rm x}$. By
 minimizing the corresponding total $\chi^2$ in Eq.~(\ref{eq17}),
 we find the best-fit parameters $\Omega_{\rm m0}=0.277$,
 $\Omega_{\rm u0}=0.214$, $d_{\rm u}=13.478$, $\alpha=0.589$,
 and $w_{\rm x}=-1.666$, while $\chi^2_{min}=311.077$. In
 Fig.~\ref{fig4}, we present the corresponding $68\%$ and
 $95\%$ C.L. contours in the $\Omega_{\rm u0}-d_{\rm u}$ plane
 and the $\alpha-d_{\rm u}$ plane for the interacting XUCDM
 model with $w_{\rm u}^{\rm eff}\not=w_{\rm x}$, while the
 other parameters are taken to be the corresponding best-fit
 values. It is fortunate to find that the unparticle dark component
 can have a sizable contribution to the total cosmic energy density
 ($0.1<\Omega_{\rm u0}<0.2$ in the $68\%$ C.L. region and
 $0.1<\Omega_{\rm u0}<0.25$ in the $95\%$ C.L. region) while
 $1\leq d_{\rm u}\leq 2$ (for bosonic unparticles) or
 $3/2\leq d_{\rm u}\leq 5/2$ (for fermionic unparticles). The
 serious situation found in~\cite{r11} could be significantly
 relaxed in this case.


\vspace{-12mm}  


\subsection{The interacting XUCDM model with
 $w_{\rm u}^{\rm eff}=w_{\rm x}$}\label{sec4c}
\begin{center}
 \begin{figure}[tbhp]
 \centering
 \includegraphics[width=0.98\textwidth]{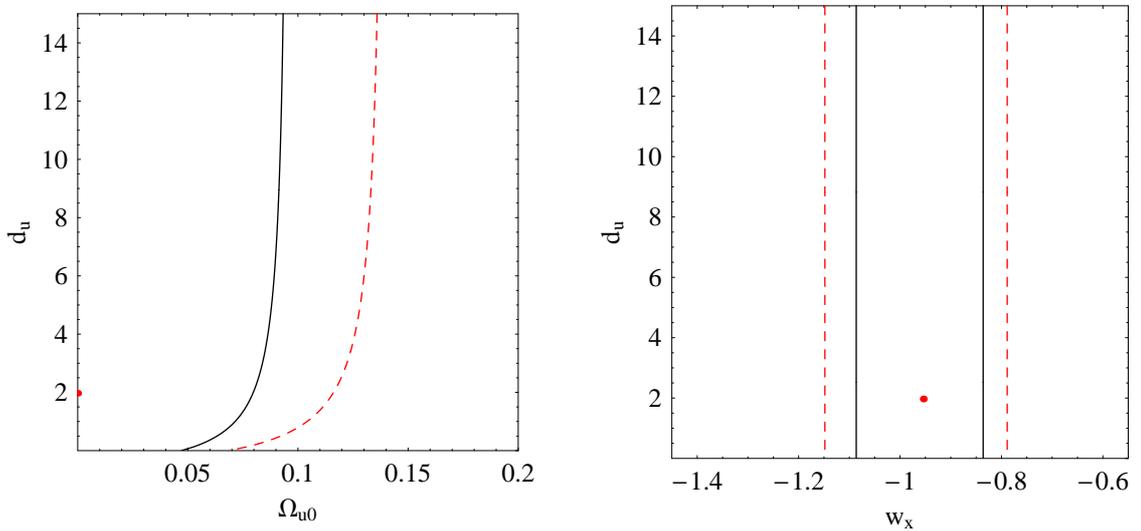}
 \caption{\label{fig5}
 The $68\%$ C.L. (black solid lines) and $95\%$ C.L. (red
 dashed lines) contours in the $\Omega_{\rm u0}-d_{\rm u}$
 plane and the $w_{\rm x}-d_{\rm u}$ plane for the interacting
 XUCDM model with $w_{\rm u}^{\rm eff}=w_{\rm x}$. The
 best fit is indicated by the red solid point.}
 \end{figure}
 \end{center}

As mentioned above, the solution Eq.~(\ref{eq26}) diverges for
 $w_{\rm u}^{\rm eff}=w_{\rm x}$. Therefore, one is required to
 consider this case separately. If $w_{\rm u}^{\rm eff}=w_{\rm x}$,
 we find that
 \be{eq28}
 \alpha=w_{\rm u}-w_{\rm x}=\frac{1}{2d_{\rm u}+1}-w_{\rm x}.
 \ee
 So, $\alpha$ is no longer a free model parameter. In this
 case, the solution for Eq.~(\ref{eq24}) is given by
 \be{eq29}
 \rho_{\rm x}=\left(const.-3\alpha\rho_{\rm u0}\ln a\right)
 a^{-3\left(1+w_{\rm x}\right)},
 \ee
 where $const.$ is an integration constant. Inserting
 Eq.~(\ref{eq29}) into the Friedmann equation and requiring
 $E(z=0)=1$, one can determine the integration constant.
 Finally, we find that
 \be{eq30}
 E^2=\Omega_{\rm m0}(1+z)^3+\left[\left(1-
 \Omega_{\rm m0}\right)+3\Omega_{\rm u0}\left(
 \frac{1}{2d_{\rm u}+1}-w_{\rm x}\right)\ln (1+z)\right]
 (1+z)^{3\left(1+w_{\rm x}\right)}.
 \ee
 There are 4 free model parameters, namely, $\Omega_{\rm m0}$,
 $\Omega_{\rm u0}$, $d_{\rm u}$ and $w_{\rm x}$. By
 minimizing the corresponding total $\chi^2$ in
 Eq.~(\ref{eq17}), we find the best-fit parameters
 $\Omega_{\rm m0}=0.270$, $\Omega_{\rm u0}=0.0$,
 $d_{\rm u}=1.969$ and $w_{\rm x}=-0.954$, while
 $\chi^2_{min}=313.138$. In Fig.~\ref{fig5}, we present the
 corresponding $68\%$ and $95\%$ C.L. contours in the
 $\Omega_{\rm u0}-d_{\rm u}$ plane and the
 $w_{\rm x}-d_{\rm u}$ plane for the interacting XUCDM
 model with $w_{\rm u}^{\rm eff}=w_{\rm x}$, while the other
 parameters are taken to be the corresponding best-fit values.
 It is a pleasure to find that the unparticle dark component
 can have a considerable contribution to the total cosmic energy
 density ($\Omega_{\rm u0}>0.1$) in the $95\%$ C.L. region
 while $1\leq d_{\rm u}\leq 2$ (for bosonic unparticles) or
 $3/2\leq d_{\rm u}\leq 5/2$ (for fermionic unparticles).
 The serious situation found in~\cite{r11} could also be
 relaxed in this case.


\vspace{-10mm}  


\section{Concluding remarks}\label{sec5}
Unparticle physics has been an active field since the seminal
 work of Georgi~\cite{r1,r2}. Recently, many constraints on
 unparticles from various observations were considered in
 the literature. In particular, the cosmological constraints
 on the unparticle dark component put it in a serious
 situation~\cite{r11}. In the present work, we try to find a
 way out of this serious situation, by including the possible
 interaction between dark energy and the unparticle dark component.
 In fact, as we have shown above, the cosmological constraints
 on the unparticle dark component can be relaxed in the interacting
 $\Lambda$UCDM model with $w_{\rm u}^{\rm eff}=-1$, and both
 the interacting XUCDM models with
 $w_{\rm u}^{\rm eff}\not=w_{\rm x}$ and
 $w_{\rm u}^{\rm eff}=w_{\rm x}$. In these interacting models,
 the unparticle dark component can have a considerable contribution
 to the total cosmic energy density (namely
 $\Omega_{\rm u0}>0.1$; even $\Omega_{\rm u0}$ can be larger
 than $0.2$ in the interacting XUCDM model with
 $w_{\rm u}^{\rm eff}\not=w_{\rm x}$) while
 $1\leq d_{\rm u}\leq 2$ (for bosonic unparticles) or
 $3/2\leq d_{\rm u}\leq 5/2$ (for fermionic unparticles),
 thanks to the possible interaction between dark energy and
 unparticle dark component. The tension between theoretical
 bounds and cosmological constraints could be removed.

It is easy to understand these results physically. When dark
 energy, the major component of the universe, can decay
 into unparticles, the energy density of the unparticle dark
 component should decrease {\em more slowly}. Since both dark
 energy and unparticle are unseen, the interaction between them
 is not prevented. If the dilution rates of unparticle dark
 component and cold dark matter are comparable, it is possible
 to have a considerable energy density of the unparticle dark
 component at the present epoch, without requiring $d_{\rm u}$
 to be very large. In fact, this is similar to the
 key point of the interacting dark energy models which are
 extensively considered in the literature to alleviate the
 cosmological coincidence problem (see e.g.~\cite{r14,r25} and
 references therein).

As shown above, the cosmological constraints on the unparticle
 dark component can be relaxed in the interacting
 $\Lambda$UCDM model with $w_{\rm u}^{\rm eff}=-1$, and
 both the interacting XUCDM models with
 $w_{\rm u}^{\rm eff}\not=w_{\rm x}$ and
 $w_{\rm u}^{\rm eff}=w_{\rm x}$. However, one might note
 that the unparticle dark component with cosmological constant
 is constrained by the observational data relatively more severe
 than in the case with dark energy. This is also easy to
 understand. Notice that the EoS of dark energy $w_{\rm x}$
 is a {\em free} model parameter, whereas the EoS of the
 cosmological constant is a constant $-1$. Therefore, the
 number of free model parameters in the interacting XUCDM model
 is larger than the one in the interacting $\Lambda$UCDM model.
 As is well known, statistically, the constraints should be
 looser in this case.


\section*{ACKNOWLEDGEMENTS}
We thank the anonymous referee for quite useful comments and
 suggestions, which helped us to improve this work. We are
 grateful to Professors Rong-Gen~Cai, Shuang~Nan~Zhang,
 Xuelei~Chen, and Xinmin~Zhang, for helpful discussions.
 We also thank Minzi~Feng, as well as Yan~Gong, Xin~Zhang and
 Dao-Jun~Liu, for kind help and discussions. We would like to
 express our gratitude to IHEP (Beijing) for kind hospitality
 during the period of National Cosmology Workshop. This work
 was supported by the Excellent Young Scholars Research
 Fund of Beijing Institute of Technology.

\renewcommand{\baselinestretch}{1.2}



\begin{thebibliography}{99}

\bibitem{r1}
H.~Georgi, Phys.\ Rev.\ Lett.\  {\bf 98}, 221601 (2007)
 [hep-ph/0703260].

\bibitem{r2}
H.~Georgi, Phys.\ Lett.\  B {\bf 650}, 275 (2007) [arXiv:0704.2457].

\bibitem{r3}
K.~Cheung, W.~Y.~Keung and T.~C.~Yuan,
 AIP Conf.\ Proc.\  {\bf 1078}, 156 (2009) [arXiv:0809.0995].

\bibitem{r4}
A.~Rajaraman,
 AIP Conf.\ Proc.\  {\bf 1078}, 63 (2009) [arXiv:0809.5092].

\bibitem{r5}
K.~Cheung, W.~Y.~Keung and T.~C.~Yuan,
 Phys.\ Rev.\ Lett.\  {\bf 99}, 051803 (2007)
 [arXiv:0704.2588];\\
K.~Cheung, W.~Y.~Keung and T.~C.~Yuan,
 Phys.\ Rev.\  D {\bf 76}, 055003 (2007)
 [arXiv:0706.3155];\\
K.~Cheung, W.~Y.~Keung and T.~C.~Yuan,
 arXiv:0710.2230 [hep-ph].

\bibitem{r6}
Y.~Liao and J.~Y.~Liu,
 Phys.\ Rev.\ Lett.\  {\bf 99}, 191804 (2007)
 [arXiv:0706.1284];\\
H.~Goldberg and P.~Nath,
 Phys.\ Rev.\ Lett.\  {\bf 100}, 031803 (2008);\\
N.~G.~Deshpande, S.~D.~H.~Hsu and J.~Jiang,
 Phys.\ Lett.\  B {\bf 659}, 888 (2008)
 [arXiv:0708.2735].

\bibitem{r7}
A.~Freitas and D.~Wyler, JHEP {\bf 0712}, 033 (2007)
 [arXiv:0708.4339].

\bibitem{r8}
M.~C.~Gonzalez-Garcia {\it et al.},
 JCAP {\bf 0806}, 019 (2008) [arXiv:0803.1180];\\
L.~Anchordoqui and H.~Goldberg,
 Phys.\ Lett.\  B {\bf 659}, 345 (2008)
 [arXiv:0709.0678].

\bibitem{r9}
H.~Davoudiasl, Phys.\ Rev.\ Lett.\  {\bf 99}, 141301 (2007)
 [arXiv:0705.3636];\\
S.~Hannestad, G.~Raffelt and Y.~Y.~Y.~Wong,
 Phys.\ Rev.\  D {\bf 76}, 121701 (2007)
 [arXiv:0708.1404];\\
P.~K.~Das, Phys.\ Rev.\  D {\bf 76}, 123012 (2007)
 [arXiv:0708.2812];\\
S.~Dutta and A.~Goyal, JCAP {\bf 0803}, 027 (2008)
 [arXiv:0712.0145].

\bibitem{r10}
J.~McDonald, arXiv:0709.2350 [hep-ph];\\
J.~McDonald, arXiv:0805.1888 [hep-ph];\\
I.~Lewis, arXiv:0710.4147 [hep-ph];\\
B.~Grzadkowski and J.~Wudka, arXiv:0809.0977 [hep-ph].

\bibitem{r11}
Y.~Gong and X.~Chen,
 Eur.\ Phys.\ J.\  C {\bf 57}, 785 (2008) [arXiv:0803.3223].

\bibitem{r12}
S.~L.~Chen, X.~G.~He, X.~P.~Hu and Y.~Liao,
 Eur.\ Phys.\ J.\  C {\bf 60}, 317 (2009) [arXiv:0710.5129].

\bibitem{r13}
N.~G.~Deshpande, X.~G.~He and J.~Jiang,
 Phys.\ Lett.\  B {\bf 656}, 91 (2007) [arXiv:0707.2959];\\
T.~Kikuchi and N.~Okada, Phys.\ Lett.\  B {\bf 665}, 186 (2008)
 [arXiv:0711.1506].

\bibitem{r14}
H.~Wei and R.~G.~Cai,
 Phys.\ Rev.\  D {\bf 71}, 043504 (2005) [hep-th/0412045];\\
H.~Wei and R.~G.~Cai,
 Phys.\ Rev.\  D {\bf 72}, 123507 (2005) [astro-ph/0509328];\\
H.~Wei and R.~G.~Cai,
 Phys.\ Rev.\  D {\bf 73}, 083002 (2006) [astro-ph/0603052];\\
H.~Wei and R.~G.~Cai,
 JCAP {\bf 0709}, 015 (2007) [astro-ph/0607064];\\
H.~Wei and S.~N.~Zhang,
 Phys.\ Lett.\  B {\bf 644}, 7 (2007) [astro-ph/0609597];\\
H.~Wei and S.~N.~Zhang,
 Phys.\ Lett.\  B {\bf 654}, 139 (2007) [arXiv:0704.3330];\\
H.~Wei and R.~G.~Cai,
 Phys.\ Lett.\  B {\bf 655}, 1 (2007) [arXiv:0707.4526];\\
H.~Wei and S.~N.~Zhang,
 Phys.\ Rev.\  D {\bf 78}, 023011 (2008) [arXiv:0803.3292];\\
H.~Wei and R.~G.~Cai,
 Eur.\ Phys.\ J.\  C {\bf 59}, 99 (2009) [arXiv:0707.4052].

\bibitem{r15}
M.~Kowalski {\it et al.},
 Astrophys.\ J.\  {\bf 686}, 749 (2008) [arXiv:0804.4142].\\
 The numerical data of the full sample are available at
 http:$/\!/$supernova.lbl.gov/Union

\bibitem{r16}
S.~Nesseris and L.~Perivolaropoulos,
 Phys.\ Rev.\ D {\bf 72}, 123519 (2005) [astro-ph/0511040];\\
L.~Perivolaropoulos,
 Phys.\ Rev.\ D {\bf 71}, 063503 (2005) [astro-ph/0412308].

\bibitem{r17}
E.~Di Pietro and J.~F.~Claeskens,
 Mon.\ Not.\ Roy.\ Astron.\ Soc.\  {\bf 341}, 1299 (2003)
 [astro-ph/0207332].

\bibitem{r18}
J.~R.~Bond, G.~Efstathiou and M.~Tegmark,
 Mon.\ Not.\ Roy.\ Astron.\ Soc.\  {\bf 291}, L33 (1997)
 [astro-ph/9702100].

\bibitem{r19}
Y.~Wang and P.~Mukherjee,
 Astrophys.\ J.\  {\bf 650}, 1 (2006) [astro-ph/0604051].

\bibitem{r20}
E.~Komatsu {\it et al.}  [WMAP Collaboration],
 Astrophys.\ J.\ Suppl.\  {\bf 180}, 330 (2009) [arXiv:0803.0547].

\bibitem{r21}
M.~Tegmark {\it et al.} [SDSS Collaboration],
 Phys.\ Rev.\ D {\bf 69}, 103501 (2004) [astro-ph/0310723];\\
M.~Tegmark {\it et al.} [SDSS Collaboration],
 Astrophys.\ J.\  {\bf 606}, 702 (2004) [astro-ph/0310725];\\
U.~Seljak {\it et al.} [SDSS Collaboration],
 Phys.\ Rev.\ D {\bf 71}, 103515 (2005) [astro-ph/0407372];\\
M.~Tegmark {\it et al.} [SDSS Collaboration],
 Phys.\ Rev.\  D {\bf 74}, 123507 (2006) [astro-ph/0608632].

\bibitem{r22}
D.~J.~Eisenstein {\it et al.} [SDSS Collaboration],
 Astrophys.\ J.\  {\bf 633}, 560 (2005) [astro-ph/0501171].

\bibitem{r23}
D.~C.~Dai and D.~Stojkovic, arXiv:0812.3396 [gr-qc].

\bibitem{r24}
F.~Sannino and R.~Zwicky,
 Phys.\ Rev.\  D {\bf 79}, 015016 (2009) [arXiv:0810.2686].

\bibitem{r25}
Z.~K.~Guo, N.~Ohta and S.~Tsujikawa,
 Phys.\ Rev.\  D {\bf 76}, 023508 (2007) [astro-ph/0702015].

\bibitem{r26}
J.~M.~Overduin and F.~I.~Cooperstock,
 Phys.\ Rev.\  D {\bf 58}, 043506 (1998) [astro-ph/9805260];\\
J.~M.~Overduin and P.~S.~Wesson,
 Phys.\ Rept.\  {\bf 402}, 267 (2004) [astro-ph/0407207].

\bibitem{r27}
P.~Wang and X.~H.~Meng,
 Class.\ Quant.\ Grav.\  {\bf 22}, 283 (2005) [astro-ph/0408495].

\bibitem{r28}
Y.~Z.~Ma,
 Nucl.\ Phys.\  B {\bf 804}, 262 (2008) [arXiv:0708.3606].

\end{thebibliography}
\end{document}